# Analysis of Metagenome Composition by the Method of Random Primers

Valery Kirzhner[1], Zeev Volkovich[2], Renata Avros[2] and Katerina Korenblat[2]

[1] Institute of Evolution, University of Haifa, Haifa 31905, Israel;
[2] Software Engineering Department, ORT Braude College of Engineering, Karmiel  21982, Israel.

**Abstract**. Metagenome, a mixture of different genomes (as a rule, bacterial), represents a pattern, and the analysis of its composition is, currently, one of the challenging problems of bioinformatics. In the present study, the possibility of evaluating metagenome composition by DNA-marker methods is investigated. These methods are based on using primers, short nucleic acid fragments. Each primer picks out of the tested genome the fragment set specific just for this genome, which is called its spectrum (for the given primer) and is used for identifying the genome. The DNA-marker method, applied to a metagenome, also gives its spectrum, which, obviously, represents the union of the spectra of all genomes belonging to the metagenome. Thus each primer provides a projection of the genomes and of the metagenome onto the corresponding spectra set. Here we propose to apply the random projection (random primer) approach for analyzing metagenome composition and present some estimates of the method effectiveness for the case of Random Amplified Polymorphic DNA (RAPD) technology.

## 1. Introduction

In the present work, we have studied the possibility of applying the random amplified polymorphic DNA (RAPD) method to analyze the metagenome composition. The method is based on cutting the genome under investigation into fragments of different lengths using a single primer - an arbitrary short sequence of nucleotides [1]. Such primers cut the genome into fragments that are random with respect to the entire set of genomes, but strictly defined for each particular genome. By the method of electrophoresis, which is, normally, part of the RAPD analysis, the resulting fragments can be (physically) distributed along a certain linear scale. In this distribution, equal-length fragments are located in the same position, the positions being different for different lengths[1] and can be considered as integer-valued. The resulting ordering appears as lines corresponding to the length of the fragments that constitute the lines and the whole picture is referred to as the genome spectrum, while the lines are called spectral lines. This distribution is an inherent genome characteristic and can be used for genome identification.

The application of the RAPD technique to a mixture of different bacteria (metagenome) will also give a spectrum, which is, obviously, the union of the spectra of the bacteria comprising the metagenome. In the present work, we study the possibility of detecting known bacteria[2] in a metagenome, using only the genome and the corresponding metagenome spectra, i.e., without the need of sequencing or other methods of distinguishing between the same-length fragments. Although our analysis is based on the simplest example of the DNA-marking methods, RAPD, the proposed approach can be transferred to other methods of the same group, for example, restriction fragment length polymorphism (RFLP) and amplified fragment length polymorphism (AFLP) (e.g. AFLP-PCR ) methods [2-4].

## 2. Results and Discussion

**2.1. Spectra of bacteria and metagenome. Algorithm for evaluating metagenome composition.**

It is convenient to represent the genome spectrum as a binary vector S, whose coordinate values of 1 or 0 indicate the presence or absence, respectively, a spectrum line in the corresponding position. Thus it can be said that spectrum S has line χ if the coordinate of χ equals 1. Let us designate the number of units in spectrum S as σ(S) and refer to this value as the *spectrum weight*. The dimensionality of vector S depends on the limitations imposed on the possible lengths of the fragments by the method of their identification. If it is possible to identify the lengths from 1 to *n*, the dimensionality of vector S equals *n*, which can be also considered as the *size of the allowable scale*.

---

[1] This distribution can sometimes depend on additional parameters of the fragments.
[2] I. e., the bacteria whose spectrum with respect to the corresponding primer is known.

The metagenome spectrum can be determined directly for each primer, but, obviously, it is the logical disjunction of the spectra (determined for the same primer) of all the bacteria comprising the metagenome. If $\boldsymbol{S^1}, \boldsymbol{S^2}, \ldots, \boldsymbol{S^p}$ are the spectra of all the bacteria that constitute metagenome *M*, spectrum $\boldsymbol{S^M}$ of this metagenome is

$$\boldsymbol{S^M} = \boldsymbol{S^1} + \boldsymbol{S^2} + \cdots + \boldsymbol{S^p} \quad , \qquad (1)$$

where «+» is logical disjunction. Thus, if one or more bacteria of the metagenome have line χ, in their spectra, this line will be present in the metagenome spectrum. On the other hand, if none of the bacteria has line χ in its spectrum, this line will be absent in the metagenome spectrum, either.

Let $\boldsymbol{S_i^M}$ be the spectrum of metagenome *M* for primer *i* and $\boldsymbol{\sigma(S_i^j \cap S_i^M)}$ be the number of non-zero coordinates common for vectors $\boldsymbol{S_i^j}$ (genome *j*) and $\boldsymbol{S_i^M}$. If

$$\boldsymbol{\sigma(S_i^j)} = \boldsymbol{\sigma(S_i^j \cap S_i^M)}, \quad \boldsymbol{\sigma(S_i^j)} \neq \boldsymbol{0}, \qquad (2)$$

then it can be said that the metagenome spectrum *covers*[3] the spectrum of genome *j* for primer *i*. We will also say that the spectrum covers the whole set of the allowable scale positions, where the spectrum lines can be located. If the spectrum covers all the allowable positions, this covering will be referred to as *total covering*.

If the spectra under consideration contain no errors, the necessary condition for a genome to belong to the metagenome is, obviously, covering, for any primer, the genome spectrum by that of the metagenome. This condition will be further used as the crucial validation of belonging a genome to the metagenome. It is obvious that such conclusion based on the necessary condition will always have a one-sided error - the erroneous recognition of certain genomes as belonging to the metagenome. However, this error can be reduced by conducting a number of tests with different primers. Indeed, for some primer $p_1$, let metagenome $M$ cover $\sigma(S^M_{p_1})$ positions and the spectrum of the genome to be tested have $\sigma(S^*_{p_1})$ lines.

---

[3] Obviously, a zero genome spectrum ( $\sigma(S_i^j) = \boldsymbol{0}$ ) is always covered by the metagenome spectrum, but such covering does not bear any information.

Provided that this genome is not included in the metagenome, the probability of covering all the lines spectrum by the metagenome spectrum is, obviously,

$$\frac{\sigma(S^M_{p_1})}{n} \frac{(\sigma(S^M_{p_1})-1)}{n-1} \ldots \frac{(\sigma(S^M_{p_1})-\sigma(S^*_{p_1})+1)}{n-\sigma(S^*_{p_1})+1} = \tau(p_1, M, *) \quad , \quad (3)$$

where где $n$ is the allowable scale dimension. The evaluation of the expression on the left side of equality (3) is based on the fact that the genome spectral lines occupy different positions. In the same way, for another primer $p_2$, the probability of covering the tested genome spectrum by the metagenome spectrum is $\tau(p_2, M, *)$. Since the covering probabilities for two different primers are independent of each other, the results of two tests give the probability of the random covering of the tested genome spectrum by the metagenome spectrum equal to $\tau(p_1, M, *) \cdot \tau(p_2, M, *)$. This process can be continued until either, for the next primer, the metagenome spectrum does not cover the genome spectrum (which means that the genome is not included in the metagenome), or the product of all the probabilities becomes less than a certain predetermined level. In the latter case it should be concluded that the genome is included in the metagenome with the latter predetermined probability.

Thus we can formulate the following

**Algorithm:**

1. **_Initialization:_** *Select set P of primers. Define the value of probability ε-level below which the result is considered to be significant. Set the value of ω=1.*

2. **_Algorithm step._** *For the next primer chosen from set P, calculate the spectrum of the tested genome and the metagenome spectrum. If the metagenome spectrum does not cover the genome spectrum, the genome does not belong to the metagenome. The* *result is obtained. If the metagenome spectrum covers the genome spectrum, multiply the value of ω by the covering probability, calculated under the condition of the genome being not included in the metagenome: $(\boldsymbol{p},\boldsymbol{M},*)$ . If ω< ε, the tested genome belongs to the metagenome with the probability larger than 1-ε. If ω> ε, execute step 2.*

Below, we analyze the effect of all the algorithm constituents on the recognition efficiency.

## 2.2. Application of the algorithm to a set of bacterial genomes.

2.2.1. <u>Bacterial genomes.</u>

In this work, the algorithm is tested using set B of 100 bacterial genomes described in ([5], Supporting Information) This set is sufficiently representative to provide the basic (typical) parameter values and the algorithm behavior when used in actual calculations.

In contrast to the metagenome under consideration, it is up to the user to choose the set of primers. Some of the average characteristics of this set can be tested on a sufficiently representative set of genomes (in our case, set B), these characteristics being only slightly different for other genome sets. To reduce the dependence of the numerical testing result on the fixed set of genomes, the primers were generated randomly, with equal probability of occupying each primer position nucleotides (below referred to as *letters*) A, T, C, G. Denote the set of 100 primers obtained in this way as P.

In this Section, the range of 50-1000 is chosen for the spectral line positions. Indeed, in 2.5 % agarose gel, the range of recognizable fragment lengths is just from 50 до 1000 [6]; thus the number of spectral line positions n =950. We assume that the spectra of each primer under consideration should have a relatively small number of lines. Since it is impossible to fix this number accurately, the above assumption will be referred to the average number of lines in a spectrum over all the genomes considered. In this paper, two groups of primers will be considered, namely, those having the average number of 10 or 30 spectral lines over the tested genome set *B*. Considering the primer length and number of allowable errors in their matching as parameters, we have chosen their values to give, on the average, 10 and 30 spectral lines over genome set B. Thus, for a random set of 100 primers of length 7, without errors, the average number of lines in the spectrum of all the bacteria which belong to set *B* is equal 9.55≈10 over the whole set of the primers considered. For a random set of 100 primer of length 11, with two errors, the same average number of lines is equal to 29.57≈30. Denote the above two primer sets as $P_{10}$ and $P_{30}$. The average number of spectral lines being fixed, the spectra of individual bacteria may differ significantly.

The calculated distribution of the groups of equal-weight spectra over set B is shown in Fig. 1. It should be noted that the scatter of the line number in the spectrum is large enough and includes spectra with zero number of lines. For the spectra that comprise set $P_{10}$, the spectra with the number of lines 0-3 account for half of all the spectra, so that the median is equal to 3, while the spectra with the number of lines more than 10 account for about 25%. For the spectra of set $P_{30}$, the median is equal to 12, while the spectra with the number of lines more than 30 also account for about 25%.

Consider metagenomes of size (the number of different bacterial genomes in a metagenome) 10 or 50. Size 10 gives the idea about the peculiarities of the analysis of metagenomes with a small number of genomes, while size 50 is, actually, enforced by the number of genomes for simulation being only 100. However, as will be shown below, the allowable number of genomes in the metagenome is relatively small for the given scale due to the resulting total covering. Therefore, metagenome of size 50 can be an example of a "large" metagenome.

Below, at the end of Section 2.2, genome set B and primer set P will be used for computer simulation of the algorithm performance.

2.2.2. The features of covering a genome spectrum by that of a metagenome.
Linear dependence of genome spectra. Standard linear dependence of the spectra vectors is enough for the existence of covering in this set of spectra. Namely, let a certain linear combination of the spectra be equal to zero:

$$a_1 S^1 + \cdots + a_p S^p + a_{p+1} S^{p+1} + \cdots + a_{p+q} S^{p+q} = 0,$$

where $0 < a_i, i = 1, 2, \ldots, p;\ 0 > a_i, i = p+1, p+2, \ldots, p+q$. Move all the terms with negative coefficients to the right-hand side of the equality:

$$a_1 S^1 + \cdots + a_p S^p = -a_{p+1} S^{p+1} - \cdots - a_{p+q} S^{p+q}.$$

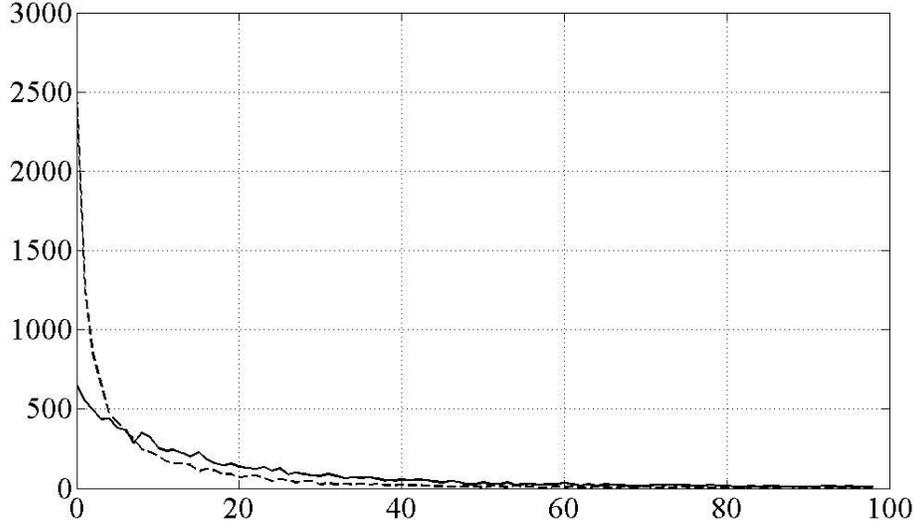

Figure 1. Distribution of the groups of equal-weight spectra over all genomes of set B for all $P_{10}$ (dashed line) and $P_{30}$ (solid line) primers. $X$ axis: the number of lines in a spectrum; $Y$ axis: the number of equal-weight spectra. The curve for $P_{10}$ primers (with the exception of point $x=0$) is approximated by $\sim \frac{x^n}{c^n + x^n}$ (Hill function) with $c=1.96$ and $n=1.26$, while the curve for $P_{30}$ primers (also with the exception of point $x=0$) is approximated by Hill function with $c=10.4$ and $n=1.6$, with correlation coefficient 0.99.

From the latter equality, it follows that in the right- and left-hand sums, the coordinate sets with zeros coincide, the same being true, obviously, for the sets with non-zero coordinates. Therefore, for the case of logical disjunction, we obtain that $S^1 + \cdots + S^p = S^{p+1} + \cdots + S^{p+q}$, i.e., that the left-hand sum covers the right-hand sum and, the more so, each term of the latter sum. Thus, *the genome spectra that are linearly dependent on the spectra of the genomes that comprise the metagenome are covered by this metagenome.* If the spectra of the genomes comprising the metagenome do not constitute a complete basis, the influence of the linear dependence can be neglected because the probability of some genome spectrum lying in the hyperplane is almost zero. However, if the metagenome contains a basis of the spectra of all the genomes under consideration, the proposed algorithm becomes invalid. On the basis of the above consideration, the following limitation on the metagenome size can be formulated: the metagenome size should not be larger than the dimension of the spectra space.

Distribution of *r* balls in *n* boxes (Occupancy problem). In fact, for the proposed algorithm being effective, the metagenome size should be much smaller than the linear dimension. The problem of covering the allowable positions of the spectral lines by the metagenome is close to the classical problem of randomly distributing r balls in n boxes. In this problem, distribution $p_m(r, n)$ of number *m* of the empty boxes is calculated. In our case, the balls are, actually, the spectral lines, while the boxes are the allowable spectral line positions, so that the value of *n* is the number of allowable positions and the value of *m* is the number of positions not covered by the metagenome spectrum.

Asymptotic of the distribution of the number of empty positions, *m*, is the Poisson distribution ([7], IV):

$$p_m(r,n) = \exp(-\lambda)\frac{\lambda^m}{m!}, \quad \lambda = n \cdot exp\left(-\frac{r}{n}\right). \quad (4)$$

where $\lambda$ is the average number of empty positions.

Our problem differs from the classical one in that the lines of one spectrum are distributed in the "boxes" not independently of each other, but, obviously, must occupy different boxes. Let us estimate the error arising if the classical model is applied to the lines of the same spectrum. According to (4), for spectra containing 10 or 30 lines, the average covering is 9.95 or 29.5, respectively, so that the error equals 0.5% or 1.7%, respectively. For a large number of spectral lines - 50 or 100 - the average covering is 49 or 95, i.e., the error is 2% or 5%, respectively.

Let us show, however, that the average error in the distribution of the spectral lines depends only on the average number of lines in the spectra set under consideration. Let $x_i$ be the fraction of spectra with weight *i* in this set. On the strength of (4), the value of $n - \lambda = n - n \cdot exp\left(-\frac{r}{n}\right)$, which the average number of covered positions under the condition of independent distribution of $r$ lines over $n$ positions. Thus the average relative error over all the spectra is:

$$\zeta = \sum_i \left( \frac{i-(n-n\cdot exp(-\frac{i}{n}))}{i} \right) x_i \quad . \quad (5)$$

Using the quadratic approximation of exponent $e^{-x} = 1 - x + x^2/2$, the average error can be estimated from (5) as:

$$\zeta \approx \sum_i \left( \frac{i}{2n} \right) x_i = \frac{\overline{\sigma(S)}}{2n}, \quad (6)$$

where $\overline{\sigma(S)}$ is the average number of the positions covered by the metagenome spectrum. For $\overline{\sigma(S)} = 9.55$ or 29.57 and n = 950, the calculated average error is 0.5% or 1.6%, respectively. Numerical verification of this result using (5) and the actual distribution of the spectra frequencies (Fig. 1) gives the same value of inaccuracy: 0.6% and 1.6% for sets $P_{10}$ and $P_{30}$, respectively. Thus, the application of the classical model of distribution introduces only minor distortions the position covering both by a single genome spectrum and by the metagenome spectrum.

Let us now estimate the value of the position covering by the metagenome spectrum. This is obviously a random value and it will be evaluated in two steps: first, the average number of lines in the metagenome will be calculate without regard to overlapping of the lines; second, the effect of overlapping will be assessed in the framework of the model.

Let $\mathbb{N}$ be a set of $N$ bacterial genomes and P be a finite set of primers. For any metagenome M, consisting of µ genomes ($|M|= µ$), and any primer $p_* \in P$, the sum of all the spectra of all the genomes belonging to M is equal to $\sum_{i \in M} \sigma(S_*^i)$. All in all, there exist $\binom{N}{\mu}$ different metagenomes consisting of µ genomes and the average weight, $\varsigma^*(µ)$, of all their spectra is

$$\varsigma^*(\mu) = \frac{1}{\binom{N}{\mu}} \sum M = \frac{1}{\binom{N}{\mu}} \sum_M \sum_{i \in M} \sigma(S_*^i) = \frac{\binom{N-1}{\mu-1}(\sigma(S_*^1)+\sigma(S_*^2)+\cdots+\sigma(S_*^N))}{\binom{N}{\mu}} = \mu \overline{\sigma(S_*)} \quad . \quad (7)$$

Next, averaging the value of $\varsigma^*(\mu)$ over all the primers of set P, we obtain

$$\overline{\sigma(\mu)} = \frac{1}{|P|} \sum_{*\in P} \varsigma^*(\mu) = \frac{1}{|P|} \sum_{*\in P} \mu \overline{\sigma(S_*)} = \mu \overline{\sigma(S)}, \quad (8)$$

where $\overline{\sigma(\mu)}$ is the average, over all the primers of set P, number of lines in a metagenome containing μ genomes and $\overline{\sigma(S)}$ is the average primer weight over all genomes of set N.

Equation (8) is a precise one. Then, according to estimation (4), the average number of empty positions is

$$\lambda = n \cdot exp\left(-\frac{\mu\overline{\sigma(S)}}{n}\right). \qquad (9)$$

Now it is possible to estimate the reasonable size of a metagenome, which can be assessed by the proposed algorithm. For $\overline{\sigma(S)}$ =9.55 the average number of positions that are not covered by the metagenome at *μ=70* or *μ=140* is, respectively, 50% и 25% of the total number of positions, *n*=950. For $\overline{\sigma(S)}$ =29.57, the same values of covering are obtained at μ=23 and μ=46. At μ=300 or μ=100, almost all 950 available positions are covered by a metagenome of such size for $\overline{\sigma(S)}$ = 10 or 30, respectively, which makes the proposed algorithm inapplicable.

2.2.3. <u>Statistical checking the algorithm and treating the output errors</u>.

To test the algorithm, the metagenomes of the two selected sizes (10 and 50 genomes), were constructed by randomly (with equal probability) choosing the required number of genomes from the 100 genomes of set B. Regardless of the metagenome choice, sets of primers of size 5 or 12 were also randomly selected from the 100 available primers of the two given above primer sets ($P_{10}$ and $P_{30}$). Then, using the tested algorithm, it was checked whether each genome of set B belonged to the current metagenome and each result was compared with the actual situation. The described procedure was repeated 1,000 times. The preciseness of the method was evaluated based on the total fraction of false-positive and false-negative errors.

Obviously, if the input data are precise, the errors may be only false-positive. Such errors can be minimized only by increasing the number of the primers being chosen. Below we will show the effectiveness of such approach by means of modeling. In what follows, we will assume that the data inaccuracy may lie in the absence of some lines in the spectrum of the metagenome under study, whereas the genome spectra taken from the existing data-base are supposed to contain no errors.

To allow for the errors, the definition of coverage should be extended. Namely, even if genome j is actually present in the metagenome, equation (2) may be transformed into inequality

$$\sigma(S_i^j) - \sigma(S_i^j \cap S_i^M) > 0 \qquad (10)$$

because part of the metagenome spectral lines may be erroneously omitted. The more so, inequality (10) can be fulfilled if genome j does not belong to the metagenome. Consider the value of difference $\sigma(S_i^j) - \sigma(S_i^j \cap S_i^M)$ with respect to the weight of the genome j spectrum:

$$\nu = \frac{\sigma(S_i^j) - \sigma(S_i^j \cap S_i^M)}{\sigma(S_i^j)} \quad , \quad \sigma(S_i^j) \neq 0 \ . \qquad (11)$$

The value of $\nu$ will be estimated by simulating 100 metagenomes with subsequent calculation of $\nu$ for 1000 random spectra. Let the error be 10%, which means that, on the average, every tenth line of the metagenome spectrum was missing.

The distribution of $\nu$ values (Fig. 2) was calculated for two situations - when genome j belongs or does not belong to the metagenome. Although the distributions overlap, the satisfactory boundary of their division can be selected as 0.3. The boundary values equal to 0.3 and less may appear for sufficiently large spectrum weights (see (11)). On the other hand, if the number of lines in the genome spectrum is small, e.g. 3, the loss of one line in the metagenome spectrum can result in $\nu = 0.33$. In order to allow for both cases, let us formulate the following rule:

*The genome spectrum is covered by the metagenome spectrum if the value of $\nu$ is not more than 0.3 or difference (10) is less than or equal to 1.*

It should be noted that this definition of coverage makes the false-negative errors also possible.

The results of simulations performed using the above-formulated rule, in the cases of accurate data and those with 10% error are shown in Table 1. It can be seen that in the case of precise data, the false-negative errors are absent, which is an obvious consequence of the algorithm. The probabilities of the false-positive errors decrease exponentially with the increase of the number of primers from 1 to 5. For example, according to Table 1, for a metagenome consisting of 10 genomes, the number of erroneously detected genomes decreases as ~$0.625^s, 1 \leq s \leq 5$, where $s$ is the number of primers from set $P_{10}$ used in testing. For a metagenome consisting of 50 genomes, the number of erroneously detected genomes decreases as ~$0.710^s, 1 \leq s \leq 5$. Similarly, for primer set $P_{30}$, the error decreases as ~$0.22^s$ and as ~$0.357^s$ ($1 \leq s \leq 5$) in the case of the metagenome size 10 and 50, respectively.

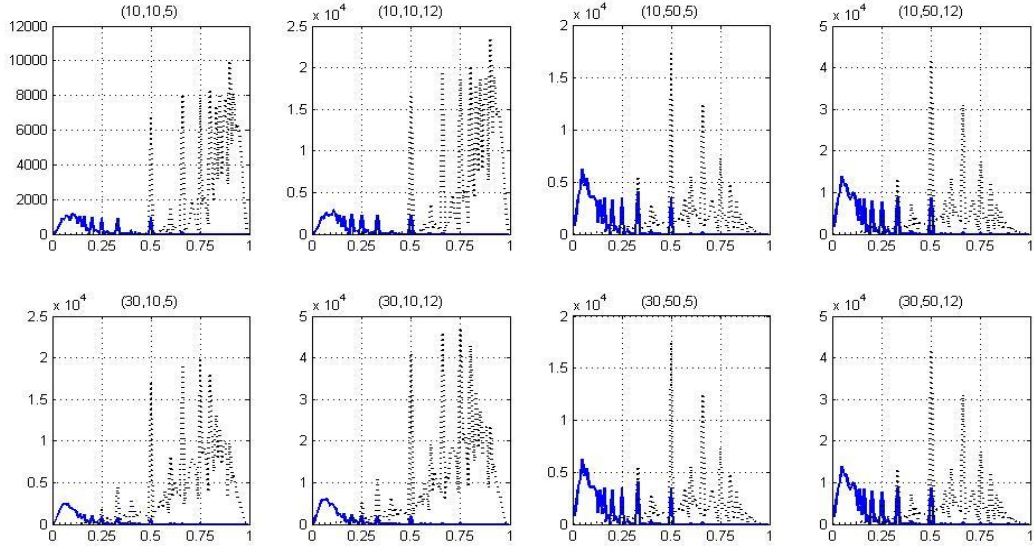

Figure 2. Distributions of $\nu$ values (11) for the cases when the genome belongs (solid line) or does not belong (dashed line) to the metagenome, at 10% error in the input metagenome spectrum. In the absence of errors, the solid line reduces to point $x=0$.

Thus, as it was expected, the rate of the error decrease is, indeed, in agreement with the assumption of independent probabilities of line overlapping. The error values as small as fractions of percent can be achieved quite easily.

Table 1. Percent of false-negative ($1 \to 0$) and false-positive ($0 \to 1$) errors for different parameters of the metagenome and the primers. The calculations were performed separately, regarding the set of genomes included or not included in the metagenome. Columns: primer weight, |S|; metagenome size |M|; the number of primers in a series, |p|. Columns ($1 \to 0$) and ($0 \to 1$): 1st column – precise input data, 2nd column – input data with a 10% error.

| |S| | |M| | |p| | | 1→0 | | 0→1 |
|---|---|---|---|---|---|---|
| 10 | 10 | 1 | 0 | 1.6 | 1.19 | 12.77 |
| 10 | 10 | 2 | 0 | 2.9 | 0.73 | 9.68 |
| 10 | 10 | 3 | 0 | 4.6 | 0.44 | 6.7 |
| 10 | 10 | 4 | 0 | 5.7 | 0.25 | 4.72 |
| 10 | 10 | 5 | 0 | 7.9 | 0.15 | 3.27 |
| 10 | 10 | 12 | 0 | 3.3 | 0.03 | 0.47 |
| 10 | 50 | 1 | 0 | 0.8 | 7.2 | 22.4 |
| 10 | 50 | 2 | 0 | 1.42 | 5.06 | 17.83 |
| 10 | 50 | 3 | 0 | 2.43 | 3.01 | 13.18 |
| 10 | 50 | 4 | 0 | 2.90 | 1.91 | 5.92 |
| 10 | 50 | 5 | 0 | 3.38 | 1.2 | 7.51 |
| 10 | 50 | 12 | 0 | 5.97 | 0.14 | 1.88 |

| |S| | |M| | |p| | | 1→0 | | 0→1 |
|---|---|---|---|---|---|---|
| 30 | 10 | 1 | 0 | 1.0 | 2.26 | 92.38 |
| 30 | 10 | 2 | 0 | 1.9 | 0.51 | 27.19 |
| 30 | 10 | 3 | 0 | 3.1 | 0.15 | 8.50 |
| 30 | 10 | 4 | 0 | 3.7 | 0.04 | 3.26 |
| 30 | 10 | 5 | 0 | 6.1 | 0.01 | 1.32 |
| 30 | 10 | 12 | 0 | 10.24 | 0 | 1 |
| 30 | 50 | 1 | 0 | 0.19 | 18.28 | 62.79 |
| 30 | 50 | 2 | 0 | 0.28 | 6.36 | 50.52 |
| 30 | 50 | 3 | 0 | 0.47 | 2.21 | 36.6 |
| 30 | 50 | 4 | 0 | 0.47 | 0.96 | 23.83 |
| 30 | 50 | 5 | 0 | 0.61 | 0.41 | 20.85 |
| 30 | 50 | 12 | 0 | 1.42 | 0.4 | 3.41 |

With a 10% error in the input metagenome spectra, misclassification may be already both false-positive and false-negative. It should be noted that the error percents on both sides are separated quite conventionally – by moving the threshold $\nu$ values (11), the error percent on one side can be increased, while the error percent on the other side is decreased. Thus the total error percent appears to be a more objective characteristic of the algorithm precision.

It can be seen from Table 1 that with the increase of the number of primers in a series from 1 to 5, the number of errors of type $(0 \to 1)$ decreases, while the number of errors of type $(1 \to 0)$ grows. The rate of type $(0 \to 1)$ error, the same as in the case of precise data, decreases exponentially, $\sim 0.78^s$, $\sim 0.79^s$ $(1 \leq s \leq 5)$, for primers belonging to set $P_{10}$ and metagenomes of size 10 and 50, respectively. For the primers belonging to set $P_{30}$ and metagenomes of size 10 and 50, the same exponential laws are $\sim 0.288^s$ **and** $\sim 0.295^s$ $(1 \leq s \leq 5)$, respectively. In contrast to the above, the number of type $(1 \to 0)$ errors grows linearly. Indeed, the increase of the number of primers in a series increases the risk of line random non-overlapping, which our rule classifies as non-belonging to the metagenome.

Above, to evaluate different algorithm constants, we have repeatedly used the simulation based on a set of genomes and primers. These constants can also be used for other input data in view of the statistical nature of the results.

2.3. Application of the algorithm to a metagenome consisting of bacterial genomes and a human genome.

Metagenomes consisting of bacterial genomes isolated from human body usually contain, in addition, a human genome. Obviously, the huge human genome is often much larger than the analyzed bacterial metagenome, so the spectral lines of the former cover a large number of spectral line positions. In this section, we analyze the possibility of applying the proposed algorithm to the study of such metagenomes, using a set of 100 random primers (words of length 10) with no more than two errors in each primer.

Designate the primer set by $H_{10}$. For each primer of set $H_{10}$, the human genome spectrum consists of a different number of spectral lines. The values of the human genome spectrum weights for different primers are shown in Fig. 3.

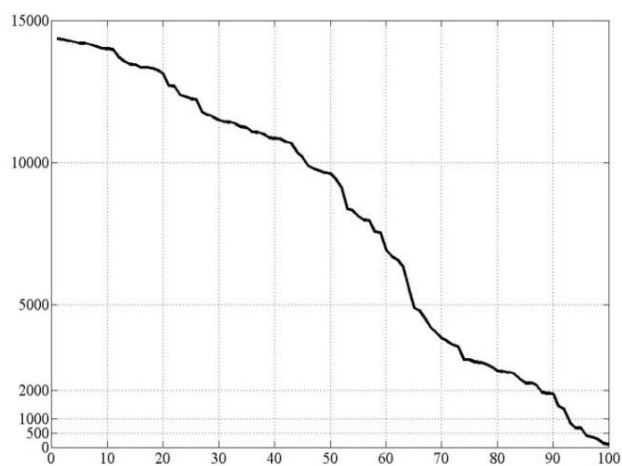

Figure 3. Human genome spectrum weights for different primers from set $H_{10}$. *X* axis: primer numbers in the order of decreasing spectrum weight; *Y* axis: spectrum weights.

Thus the number of positions covered by a human genome spectrum can be much larger that the range of 50-1000 used in the previous simulations. Therefore, it is necessary to expand the number of possible spectral line positions. Since the same mixture of fragments can be separated on gels of various concentrations, the number of the spectral lines can be increased from 50 to15,000 [6], which makes it possible to apply the proposed algorithm also to this case.

According to our calculations, the largest value of the human genome spectrum weight is equal to 14362, which leaves 600 non-occupied positions. For the bacteria from set B, the $H_{10}$ system of primers gives, on average, 17 spectral lines. The distribution of the primer weights over all bacteria of set B is presented in Fig. 4.

Let us now consider a simulation similar to the one performed in the previous section. Namely, for metagenomes consisting of 10 and 50 bacterial genomes, with a human genome included, consider series of 5 and 12 primers. Each configuration of a metagenome and a primer series is chosen randomly 1,000 times. Using the proposed algorithm, the metagenome composition in each sample was evaluated and compared with the actual one. The algorithm parameters were the same as before, for the case of a metagenome without a human genome. The simulation results are shown in Table 2.

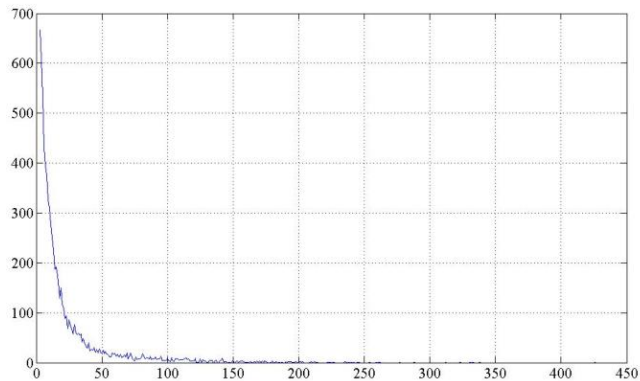

Figure 4. Distribution of the spectra weights for all primers from set $H_{10}$ and all bacteria from set B. *X* axis: spectrum weight; *Y* axis: the number of spectra with such weight, beginning from weight 2. The total number of spectra with zero and unit weight is 1229 and 911, respectively.

| |M| | |p| | | $(1 \rightarrow 0)$ (%) | | $(0 \rightarrow 1)$ (%) |
|---|---|---|---|---|---|
| 10 | 5  | 0 | 10.0 | 1.5 | 12.4 |
| 10 | 12 | 0 | 10.0 | 0.0 | 6.0  |
| 50 | 5  | 0 | 10.0 | 1.2 | 13.5 |
| 50 | 12 | 0 | 2.0  | 0.0 | 1.6  |

Table 2. Percent of false-negative $(1 \rightarrow 0)$ and false-positive $(0 \rightarrow 1)$ errors for different parameters of the metagenome and the primers. Designations are the same as in Table 1.

It can be seen from Table 2 that the errors in the case of precise data are quite small, especially for a sufficiently large series of primers. In the case of data with 10% error, there may also be 10% of erroneous results.

3. Conclusion

The proposed algorithm is intended to determine whether a single genome belongs to the metagenome under study. The composition of the metagenome is not limited in any way; in particular, it may contain unknown genomes. In most of the examples discussed above, the number of different genomes in the metagenome was limited to 50 because the number of possible spectral line positions was only 1,000. However, in the example with the human genome, 15,000 possible positions were used. It can be shown that in this case, a bacterial metagenome composed of up to 3,000 different genomes can be considered, under the condition of the average weight of the primer spectra being 10. According to relation (9), with such average weight, 15% the spectral positions remain uncovered.

Some estimates of the effectiveness of the proposed algorithm were done in the present work. For this, series of spectra obtained on the basis of random primer series were considered. It was the primer randomness and, consequently, the spectra randomness, that allowed obtaining non-covering of the tested bacterium spectrum by at least one metagenome spectrum. For accurate input experimental data, the estimation of the proposed method effectiveness is a statistical problem, which has an exact solution. The algorithm can provide the result with any pre-set accuracy.

For the input data containing errors, the result of the algorithm performance is not so obvious; however, the above simulations show the algorithm error can be estimated. In this case, the success is based on the initial non-coincidence of the bacteria spectra with respect to numerous lines because the coincidence would be random. The situation when the spectra of two randomly chosen bacteria differ by no more than 10% appears to be a rare event[4]. Therefore, the spectrum with the error as big as 10% is close to the true one.

The set of random primers used by us is, certainly, not optimized, one reason being the existence of numerous empty spectra for the genomes that we are looking for in the metagenome. Obviously, this fact greatly decreases the result quality. However, it should be noted that the primer set has not been optimized in order to represent, so to say, the "lower bound" of the algorithm effectiveness.

In the algorithm adaptation, we used the data on the primer set that the user actually possesses. In this regard, it was shown above that some important parameters depend not on each primer, but on the average number of the spectral lines. Other algorithm parameters, in particular, those used for the estimation of covering in the case of erroneous data, were obtained for a sufficiently representative genome set and are recommended to be used directly. At the same time, the user can obtain these estimates by himself for the genome and primer sets more suitable in his case, in the same way as it was done in the present study.

---

[4] If the bacteria are not classified as close. In the opposite case, replacement of the true result for a close bacterium is quite allowable.

The algorithm proposed in this work can be used with the DNA-marker method of any type.

References

1. Williams JG, Kubelik AR, Livak KJ, Rafalski JA, Tingey SV (1990). «DNA polymorphisms amplified by arbitrary primers are useful as genetic markers». *Nucleic Acids Research* 18 (22): 6531-6535.

2. Saiki, RK; Scharf S; Faloona F; Mullis KB; Erlich HA; Arnheim N (1985). "Enzymatic amplification of beta-globin genomic sequences and restriction site analysis for diagnosis of sickle cell anemia". Science. 230 (4732): 1350–1354

2. Mueller UG, Wolfenbarger LL ( 1999). "AFLP genotyping and fingerprinting". Trends Ecol. Evol. (Amst.). 14 (10): 389–394.

4. Vos P, Hogers R, Bleeker M; et al. ( 1995). "AFLP: a new technique for DNA fingerprinting". Nucleic Acids Res. 23 (21): 4407–14.

5. Kirzhner, V.; Volkovich, Z. (2012). "Evaluation of the Genome Mixture Contents by Means of the Compositional Spectra Method". Eprint Arxiv: 1203.2178, 03/2012

6. Sambrook, J., Fritsch, E.F. and Maniatis, T. (1989). Gel electrophoresis of DNA. In: Sambrook, J., Fritsch, E.F. and Maniatis, T. (Eds.)  Molecular Cloning: a Laboratory Manual. New York: Cold Spring Harbor Laboratory Press, Cold Spring Harbor, NY, USA, chapter 5.

7. W. Feller  An Introduction to Probability Theory and Its Applications. Vol. 1. 2nd. Ed. John Wiley, 1971.